\documentstyle[preprint,aps,epsfig,pre]{revtex}
\begin{document}
\draft

\title{Limit cycle induced by multiplicative noise in a system of
coupled Brownian motors}

\author{S. E. Mangioni$^a$ \thanks{E-mail: smangio@mdp.edu.ar}
and H. S. Wio$^b$ \thanks{E-mail: wio@cab.cnea.gov.ar,
wio@imedea.uib.es}}

\address{(a) Departamento de F\'{\i}sica, FCEyN, Universidad
Nacional de Mar del Plata\\ De\'an Funes 3350, 7600 Mar del Plata,
Argentina. \\ (b) Grupo de F\'{\i}sica Estad\'{\i}stica,
Centro At\'omico Bariloche (CNEA) and \\ Instituto Balseiro (CNEA
and UNCuyo), 8400 San Carlos de Bariloche, Argentina.}

\date{\today}
\maketitle

\begin{abstract}
We study a model consisting of $N$ nonlinear oscillators with {\em
global periodic\/} coupling and {\em local multiplicative\/} and
additive noises. The model was shown to undergo a nonequilibrium
phase transition towards a broken-symmetry phase exhibiting
noise-induced ``ratchet" behavior. A previous study \cite{[7]}
focused on the relationship between the character of the
hysteresis loop, the number of ``homogeneous'' mean-field
solutions and the shape of the stationary mean-field probability
distribution function. Here we show --as suggested by the absence
of stable solutions when the load force is beyond a critical
value-- the existence of a limit cycle induced by both:
multiplicative noise and {\em global periodic\/} coupling.
\end{abstract}


\section{Introduction}\label{sec.1}

The study of dynamical systems has shown that limit cycles are
ubiquitous in a wide range of physical applications
\cite{[01],[02]}. From a physicist's point of view, limit cycles
are thought of as a way to balance the in- and out- energy flows.
Even when those flows are not oscillatory in time, a system's
oscillatory motion can occur equalizing such flows over one
period. A nice pedagogical example of such a process, based on a
perturbative analysis of the nonlinear van der Pol oscillator, can
be found in \cite{[03]}. As is well known, limit cycles are robust
--structurally stable under small perturbations-- attractors in
dissipative systems without external oscillations
\cite{[01],[02]}. Usually, limit cycles arise in dynamical systems
described by ordinary differential equations (ODE)
\cite{[01],[02]}, but there are several examples where such kind
of cycles also arise in partial differential equations (PDE) or
``extended systems", as for instance, in the ''brusselator" model
for the so called ``chemical clocks" \cite{[04],[05]}.

Limit cycles arise also in systems with noise. Noise or
fluctuations, that are present everywhere, have been generally
considered as a factor that destroys order. However, a wealth of
investigations on nonlinear physics during the last decades have
shown numerous examples, both in zero- and higher dimensional
systems, of nonequilibrium systems where noise plays an
``ordering" role. In such cases, the transfer of concepts from
equilibrium thermodynamics, in order to study phenomena away from
equilibrium, is not always adequate and many times is misleading.
Some examples of such nonequilibrium phenomena are: noise induced
unimodal-bimodal transitions in some zero dimensional models
(describing either concentrated systems or uniforms fields)
\cite{[3]}, shifts in critical points \cite{[4]}, stochastic
resonance in zero-dimensional and extended systems \cite{[5],[6]},
noise-delayed decay of unstable states \cite{[7]}, noise-induced
spatial patterns \cite{[8]}, noise induced phase transitions in
extended systems \cite{[9]}, etc.

Here, we discuss an extended system described by PDE's, where
noise plays a key role in controlling and inducing a limit cycle.
The model that we analyze here is the one used in \cite{[10],[11]}
to study a ratchet-like transport mechanism arising through a
symmetry breaking, noise--induced, nonequilibrium phase
transition. In a recent paper \cite{[10p]} a system with a noise
induced phase transition, based on a model that is a variant of
Kuramoto's model for coupled phase oscillators \cite{kura}; was
analyzed. In addition to the phenomenon of anomalous hysteresis,
evidence of the existence of a limit cycle for a given parameter
region; is also shown.

The model we analyze consists of a system of periodically coupled
nonlinear phase oscillators with a multiplicative white noise.
Coupled oscillators have been used to model systems with
collective dynamics exhibiting plenty of interesting properties
like equilibrium and nonequilibrium phase transitions, coherence,
synchronization, segregation and clustering phenomena. In this
particular model a ratchet-like transport mechanism arises through
a symmetry breaking, noise--induced, nonequilibrium phase
transition \cite{[10]}, produced by the simultaneous effect of
coupling between the oscillators and the presence of a
multiplicative noise. The symmetry breaking does not arise in the
absence of any of these two ingredients. In \cite{[10]} it was
also shown that the current, as a function of a load force $F$,
produces an anomalous (clockwise) hysteresis cycle. Recently we
have reported that, changing the multiplicative noise intensity
$Q$ and/or the coupled constant $K_0$, a transition from anomalous
to normal (counter-clockwise) hysteresis is produced \cite{[11]}.
The result was obtained exploiting a mean field approximation. The
transition curve in the plane $(K_0, Q)$, separating the region
where the hysteresis cycle is anomalous from the one where it is
normal, was clearly determined.

Here we focus on the time behavior. We use a method for detecting
the existence of a limit cycle based on the evaluation of the
distance between two solutions separated by a (fixed) time
interval \cite{[12]}. In this way, we not only show the existence
of a limit cycle for $F > F_c$ (with $F_c$ a loading threshold
value), but also determine its period. We also found the time
dependence of the probability distribution function along the
cycle and calculate the order parameter of the model vs. $t$,
clearly showing the limit cycle. Next, we gain insight into its
origin through the study of the large coupling limit ($K_0 \to
\infty$). Finally, we draw some conclusions.

\section{The model, mean field and the method used }\label{sec.2}

For completeness we present a brief description of our model,
which is similar to the one used in Refs. \cite{[10]} and
\cite{[11]}. We consider a set of globally coupled stochastic
differential equations (to be interpreted in the sense of
Stratonovich) for N degrees of freedom (phases) $X_i(t)$
\begin{equation}
\dot X_i=-\frac{\partial U_i}{\partial X_i}+\sqrt{2T}\,
\xi_i(t)-\frac{1}{N}\sum_{j=1}^N K(X_i-X_j).\label{eq:1}
\end{equation}
This model can be visualized (at least for some parameter values)
as a set of overdamped interacting pendulums. The second term in
Eq. (1) considers the effect of thermal fluctuations: $T$ is the
temperature of the environment and the $\xi_i(t)$ are additive
Gaussian white noises with
\begin{equation}
\langle\xi_i(t)\rangle=0,\qquad\langle\xi_i(t)\xi_j(t')\rangle=
\delta_{ij}\delta(t-t').\label{eq:2}
\end{equation}
The last term in Eq. (1) represents the interaction force between
the oscillators. It is assumed  to fulfill $K(x-y)=-K(y-x)$ and to
be a periodic function of $x-y$ with period $L=2\,\pi$. We adopt
\cite{[10],[11]}
\begin{equation}
K(x)=K_0\sin x,\qquad K_0>0.\label{eq:3}
\end{equation}
The potential $U_i(x,t)$ consists in a static part $V(x)$ and a
fluctuating one. Gaussian white noises  $\eta_i(t)$, with zero
mean and variance 1, are introduced in a multiplicative way (with
intensity $Q$) through a function $W(x)$. In addition; a load
force $F$, producing an additional bias, is considered
\begin{equation}
U_i(x,t)=V(x)+W(x)\sqrt{2Q}\,\eta_i(t)-Fx.\label{eq:4}
\end{equation}

In addition to the interaction $K(x-y)$, $V(x)$ and $W(x)$ are
also assumed to be periodic and, furthermore, to be symmetric:
$V(x) = V(-x)$ and $W(x) = W(-x)$. This last aspect indicates that
there is no built-in ratchet effect. The form we choose is
\cite{[10],[11]}
\begin{equation}
V(x)=W(x)=-\cos x-A\cos{2x}.\label{eq:5}
\end{equation}

We introduce a mean-field approximation (MFA) similar to the one
used in Ref. \cite{[11]}. The interparticle interaction term in
Eq. (1) can be cast in the form
\begin{equation}
\frac{1}{N}\sum_{j=1}^N K(X_i-X_j)= K_0\left[C_i(t)\sin
X_i-S_i(t)\cos X_i\right].\label{eq:6}
\end{equation}
For $N \to \infty$, we may approximate Eq.\ (\ref{eq:6}) in the
Curie-Weiss form, replacing $C_i(t)\equiv N^{-1}\sum_j\cos X_j(t)$
and $S_i(t)\equiv N^{-1} \sum_j\sin X_j(t)$ by $C_m \equiv \langle
\cos X_j \rangle$ and $S_m \equiv \langle \sin X_j \rangle$,
respectively. As usual, both $C_m$ and $S_m$ should be determined
by self-consistency. This decouples the system of stochastic
differential equations (SDE) in Eq.\ (\ref{eq:1}) \, which reduces
to essentially one Markovian SDE for the single stochastic process
$X(t)$
\begin{equation}
\dot X=R(X)+S(X)\eta(t),\label{eq:7}
\end{equation}
with (hereafter, the primes will indicate derivatives with respect
to $x$)
\begin{eqnarray}
R(x)&=& -V'(x)+F-K_m(x)\nonumber\\
&=&-\sin x(1+K_0C_m+4A\cos x)+K_0S_m\cos x+F,\label{eq:8}
\end{eqnarray}
(where $K_m(x)=K_0[C_m\sin x-S_m\cos x]$) and
\begin{equation}
S(x)=\sqrt{2\{T+Q[W'(x)]^2\}}=\sqrt{2\{T+Q[\sin x+2A\sin 2x]^2\}}.
\label{eq:9}
\end{equation}

The Fokker-Planck equation (FPE) associated with the SDE in Eq.
(\ref{eq:7}) (in Stratonovich's sense) is
\begin{equation}
\partial_t P(x,t)=\partial_x \left( -[R(x)+\frac{1}{2}S(x)\,S'(x)]P(x,t)
\right) +\frac{1}{2}\partial_{xx} \left[ S^2(x)P(x,t) \right]
\label{eq:10}
\end{equation}
where $P(x,t)$ is the probability distribution function (PDF).

In \cite{[11]} we have shown that in the so called ''interaction
driven regime" (IDR) --where the hysteretic cycle is  anomalous--
and for each $F$ value, in addition to the two stationary stable
solutions with the corresponding values of current there are other
three unstable ones. Two of them merge with the two stable,
yielding a closed curve of current vs. $F$. Beyond a critical
(absolute) value of the load force $F$, indicated by $F_c$, those
stable solutions disappear. This does not happen for the ''noise
driven regime" (NDR) --where the hysteretic cycle is normal--,
where for each $F$ value, one stationary stable solution exists
(for small $|F|$ even two stationary stable solutions and an
unstable one exist).

It is worth remarking here that the absence of a stationary stable
solution, beyond the critical value $F_c$ in the IDR, suggest the
possibility that a limit cycle exists. Already in \cite{[10]}, in
a strong coupling analysis (that is considering the limit $K_0 \to
\infty$), it was indicated that for very large $|F|$ the
probability distribution function approaches a periodic long time
behavior.

In order to analyze the existence of a limit cycle, we exploit a
novel method used in Ref. \cite{[12]}. It is based on the
measurement of the distance between different solutions of a
system and evaluating its evolution in time. The approach applied
in Ref. \cite{[12]} uses a generalization of the known
Kullback-Leibler information function \cite{[13]}, which is based
on the nonextensive thermostatistics proposed by Tsallis
\cite{[14]}. Within such a formalism, the exponential and
logarithmic functions are generalized according to the following
definitions \cite{[12]}
\begin{eqnarray}
\exp_q(x)&=&[1 + (1 - q)x]^{1/(1-q)}\nonumber \\
\ln_q(x)&=&\frac{x^{1-q}-1}{1-q}.\label{eq:11}
\end{eqnarray}
The distance can be measured between an evolved initial condition
and a known stable stationary solution, or between two solutions
at different times (separated by a time interval $\Delta\tau$
which is fixed along the whole calculation). In this work we
choose the later. In Ref. \cite{[12]} the following definition for
the distance between two solutions of a reaction-diffusion
equation was adopted (valid for both indicated criteria)
\begin{equation}
I_q(P_{t+\Delta\tau}, P_{t})=-\int P_{t+ \Delta \tau}(x,t+ \Delta
\tau) \ln_q\left[\frac{P_{\tau}(x,t)} {P_{t+ \Delta \tau}(x,t+
\Delta \tau)}\right]dx,\label{eq:12}
\end{equation}
where $P$ represent a (probability-like) distribution (necessary
to use the information theory formalism), evaluated at $t$ and $t+
\Delta \tau$, according to the criterion that we have chosen. We
used this definition of distance, and evaluated
$I_q(P_{t+\Delta\tau},P_{t})$, using for $P$ the PDF obtained
solving the FPE Eq. (\ref{eq:10}). We adopted $q=2$, as it is the
value for which the sensibility of the method seems to be a
maximum \cite{[12]}. The FPE was numerically solved with a
Runge-Kuta method, using a time step $\delta t=6.25\, 10^{-7}$ and
a space interval $\delta x=0.02944$. We have tested that
variations in both steps, $\delta t$ and $\delta x$, produce no
changes in our results. Remembering that $C_m $ and $S_m$ should
be determined self-consistently, at each time step both were
calculated with the modified PDF. As our initial condition we
adopted one stationary solution for $F < F_c$ calculated as in
Ref. \cite{[11]}. The integral in Eq.\ (\ref{eq:12}) was
calculated simultaneously. Furthermore, we also obtained $v_m$
--the particle mean velocity--
\begin{equation}
v_m=\langle\dot X\rangle= \int_{-L/2}^{L/2}\, dx
\left[R(x)+\frac{1}{2}S(x)S'(x)\right] P^{st}(x,C_m, S_m),
\label{eq:13}
\end{equation}
which is adopted as the order parameter like in \cite{[11]}.

\section{Results }

\subsection{ Numerical results }\label{sec.3}

Figure 1 shows $I_q(P_{t+\Delta\tau}, P_{t})$ (normalized to its
maximum) vs. $t$ for  $A=0.15$, $T=2$, $K_0=10$, $Q=3$ and $F=1.5$
(a set of parameters for which a stationary stable solution does
not exists: see Fig 6 in Ref. \cite{[11]}). We observe that $I_q$
is a periodic function of time. This form seems to be typical for
limit cycles as shown in Ref. \cite{[12]}, the period
corresponding to the distance between peaks. In Fig. 2, for the
same parameter values, we depict the PDF at different times along
the complete cycle, where the behavior resembles a wave. In Fig. 3
we show $v_m$ and $S_m$ vs. $t$. They have a time periodic
behavior, not as in the case $F \leq F_c$, where $v_m$ ($\neq 0$)
and $S_m$ ($\neq 0$) are both constants in time. We have also
verified that the transition to the limit cycle occurs just at
$F_c$ (in this case $F_c=1.2$).

\subsection{Asymptotic strong coupling analysis }\label{sec.4}

In order to understand the origin of the previous results and gain
some insight about them, we have performed an asymptotic strong
coupling analysis. That is, we consider $K_0 \to \infty$, $P \to
\delta (x-x_m)$, hence Eq.\ (\ref{eq:13}) transforms into
\begin{equation}
\dot x_m=R(x_m)+\frac{1}{2}S(x_m)S'(x_m).\label{eq:14}
\end{equation}
A simple calculation shows
\begin{equation}
\dot x_m =- \sin x_m [1+4A \cos x_m][1-Q \cos x_m - 4AQ (1-2
\sin^2 x_m )] + F.\label{eq:15}
\end{equation}
This equation can be analyzed considering an effective potential
$U(x_m)$, given by
\begin{equation}
U(x_m) = V(x_m) - Q W'^2(x_m)/2 - F x_m, \label{eq:16}
\end{equation}
that allows us to rewrite Eq. (14) as
\begin{equation}
\dot x_m = - \frac{\partial U(x_m)}{\partial x}.\label{eq:17}
\end{equation}
Figure 4 shows the solution of Eq. (\ref{eq:15}), $\dot x_m$ vs.
$t$, for both situations: just below and above $F_c$. It was
observed that while for $F<F_c$, after a transient, the solution
becomes stationary, for $F>F_c$ it becomes oscillatory. In the
first case $x_m$ is constant in time but it does not imply $v_m=0$
because, it should be calculated with $S_m = \sin (x_m) \neq 0$,
not as in the case with $\dot x_m$. Figure 5 shows the effective
potential $U$ vs. $x_m$ for the same cases, and also for $F=0$. It
is apparent that in the first case ($F<F_c$) the potential has
only one minimum while for the second one, both possible minima
are washed out. The latter happens just when the transition to the
oscillating regime occurs. It is worth remarking here that, if
$K_0 \to \infty$, the hysteresis cycle is anomalous and closed,
and a critical load force establishing a threshold for a limit
cycle transition always exists.

\section{Conclusions }\label{sec.5}

A wealth of papers have reported on research where, by changing a
control parameter, a transition to a limit cycle occurs
\cite{[18]}. However, studies on the existence of limit cycles
under (or induced by) the influence of noise are scarce
\cite{[10p],[19],[20]}. Such an aspect was analyzed here, where we
have studied a system of periodically coupled nonlinear
oscillators  with multiplicative white noises, yielding a
ratchet-like transport mechanism through a symmetry-breaking,
noise--induced, nonequilibrium phase transition \cite{[10],[11]}.
The model includes a load force $F$, used as a control parameter,
so that the picture of the current vs. $F$ shows hysteretic
behavior.

In \cite{[11]} we have found that in the IDR the cycle is
anomalous, yielding a closed curve current vs. $F$ when the
stationary stable solutions merge with two of the three unstable
ones. For $F>F_c$ (force value at which a stable solution merges
with an unstable one) there are no stationary stable solutions.
Here we have shown, by analyzing the time evolution of the
distance between different solutions, that at $F=F_c$ a transition
to a limit cycle occurs. Such a distance shows, for $F>F_c$, a
typical periodic behavior evidencing a limit cycle \cite{[12]}.
Focusing on the analysis of the time behavior, we have shown the
evolution of both the PDF and the current, showing in both cases
the time periodicity (a time evolution of the PDF resembling a
wave). In order to understand the origin of this transition, we
have made a "strong coupling" limit analysis. It indicates that
the minima of the effective potential are "washed out" as $F$ is
increased and all the stationary stable solution are removed with
them.

As indicated in the introduction, limit cycles balance the in--
and out-- energy flows --even when those flows are not oscillatory
in time-- through a system's oscillatory motion that equalize such
flows over one period. In the present case we have found a limit
cycle in a dynamical system described by PDE's, where the energy
inflow is provided by both the load force $F$ and the noise terms,
while energy is lost (as the system is an overdamped one)
proportionally to the particle's velocity. A remarkable aspect is
the fact that it is the multiplicative noise intensity the
parameter controlling the bifurcation towards the limit cycle.

Summarizing, for this model (that is just one example among many
possible others) we have found a transition towards a limit cycle
induced by both, a multiplicative noise and a {\em global
periodic\/} coupling. However, when the noise or coupling are not
present, such a transition does not happen. This is a new feature
of those systems showing a ratchet-like transport mechanism
arising through a symmetry-breaking, noise--induced,
nonequilibrium phase transition. Also, it is another example where
the presence of a multiplicative noise contributes to build up
some form of order. \\

\noindent{\small{\bf ACKNOWLEDGMENTS:}} The authors thank
M.Hoyuelos and M.G. Wio for revision of the manuscript. Partial
support from ANPCyT, Argentina, is greatly acknowledged.

\newpage

FIGURE CAPTIONS \\

{\bf Figure 1}: $I_q(P_{t+\Delta\tau},P_{t})$ (divided by its
maximum) vs. time ($t$) for  $A=0.15$, $T=2$, $K_0=10$, $Q=3$ and
$F=1.5$ (for this set of parameters there is no stationary stable
solution).

{\bf Figure 2}: PDF ($P$) vs. $x$ for different time ($t$)
following the complete cycle (starting at $t=2.83$, and evaluated
each $\Delta \tau=0.2444$).  The parameters $A, T, K_0, Q$ and $F$
as in Fig. 1.

{\bf Figure 3}: $v_m$ and $S_m$ vs. time ($t$). The parameters $A,
T, K_0, Q$ and $F$ as in Fig. 1. Line thick for $v_m$ and thin for
$S_m$.

{\bf Figure 4}: Solution of Eq.\ (\ref{eq:15}) \, $\dot x_m$ vs.
$t$, for both situations, just below and above $F_c=1.2$. The
parameters $A, T, K_0$ and  $Q$ as in Fig. 1.  We observe that
while for $F<F_c$, after a transient, the solution becomes
stationary, for $F>F_c$ it is oscillatory. The parameters are
$K_0=10$ and $Q=3$. The solid line indicates the case $F>F_c$ and
the dashed one the case $F<F_c$.

{\bf Figure 5}: $U$ vs. $x_m$ just below and above of $F_c=1.2$.
Also the case $F=0$ is shown. The parameters are $A=0.15$,
$K_0=10$, and $Q=3$. It is observed that in the first case
($F<F_c$) the potential has at least a minimum, while for the
second one both possible minima are washed out. The solid line
indicates the case just above $F_c$ ($F_c=1.2$), the dotted
indicates the case $ F< F_c$ and dashed one the case $F=0$.

\newpage

\begin{figure}[h]
\centering
\resizebox{.5\columnwidth}{!}{\rotatebox{-90}{\includegraphics{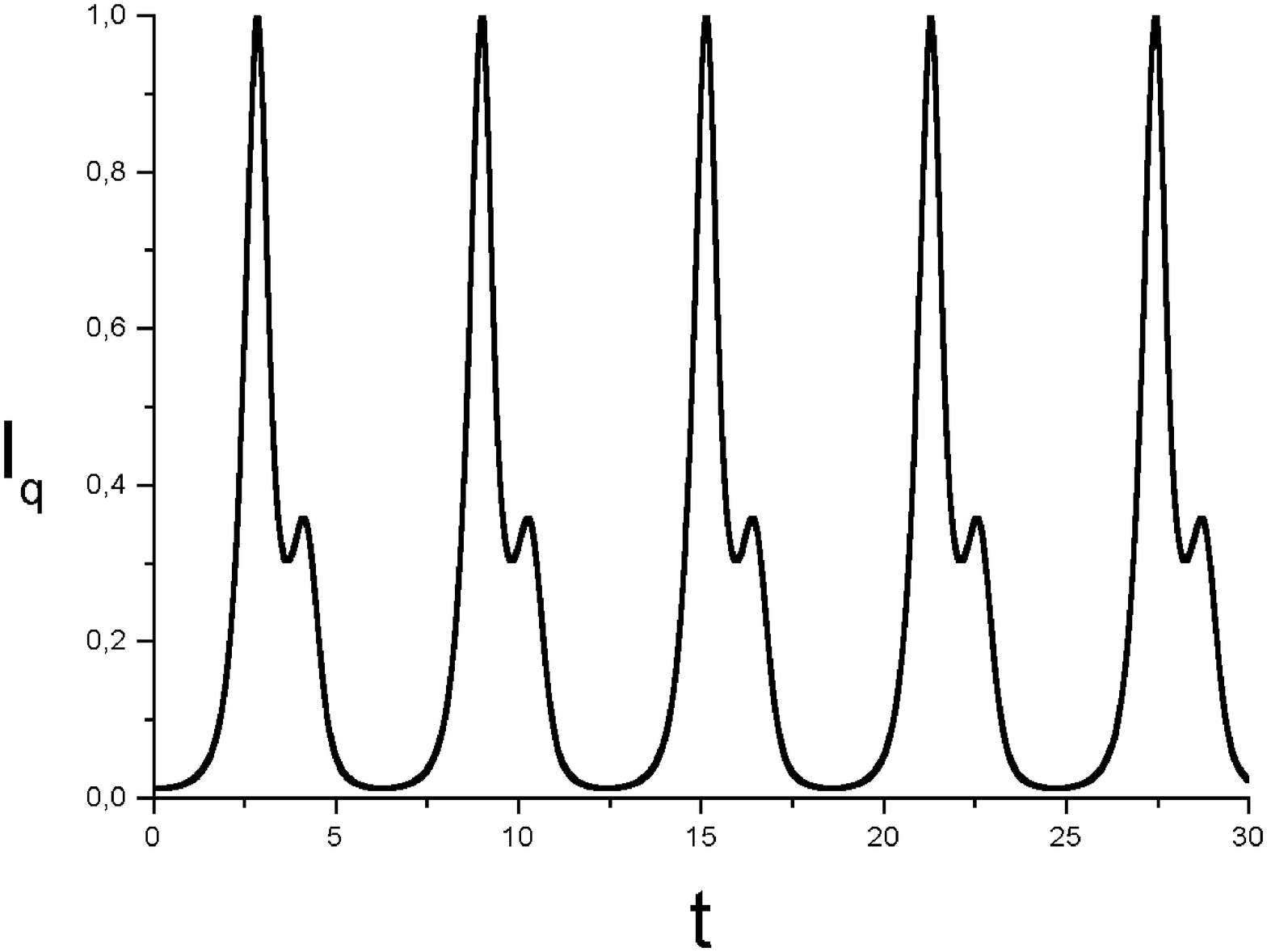}}}
\caption{} \label{fig:1}
\end{figure}

\newpage
\begin{figure}[h]
\centering
\resizebox{.6\columnwidth}{!}{\includegraphics{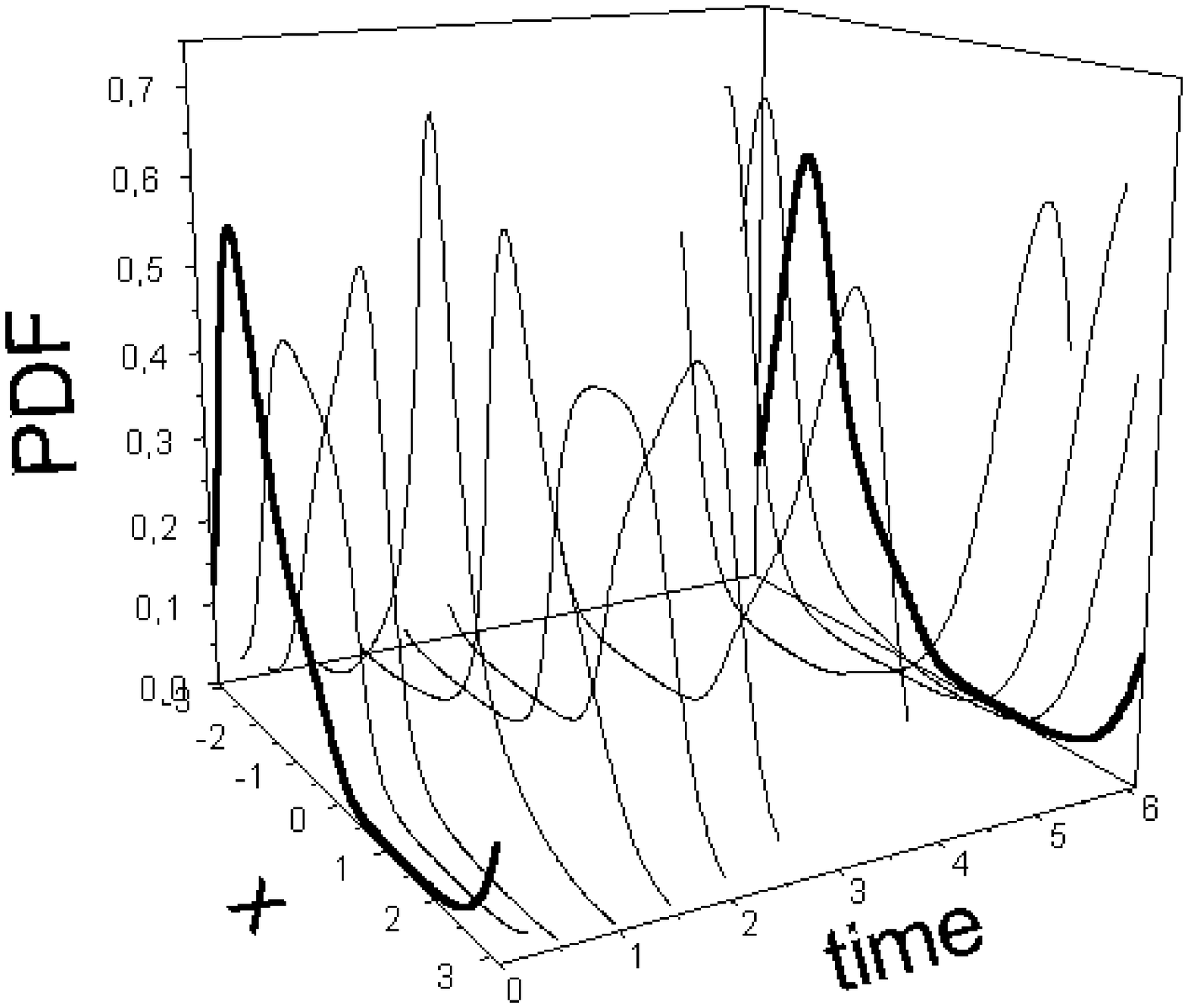}}
\caption{} \label{fig:2}
\end{figure}

\newpage

\begin{figure}[h]
\centering
\resizebox{.5\columnwidth}{!}{\rotatebox{-90}{\includegraphics{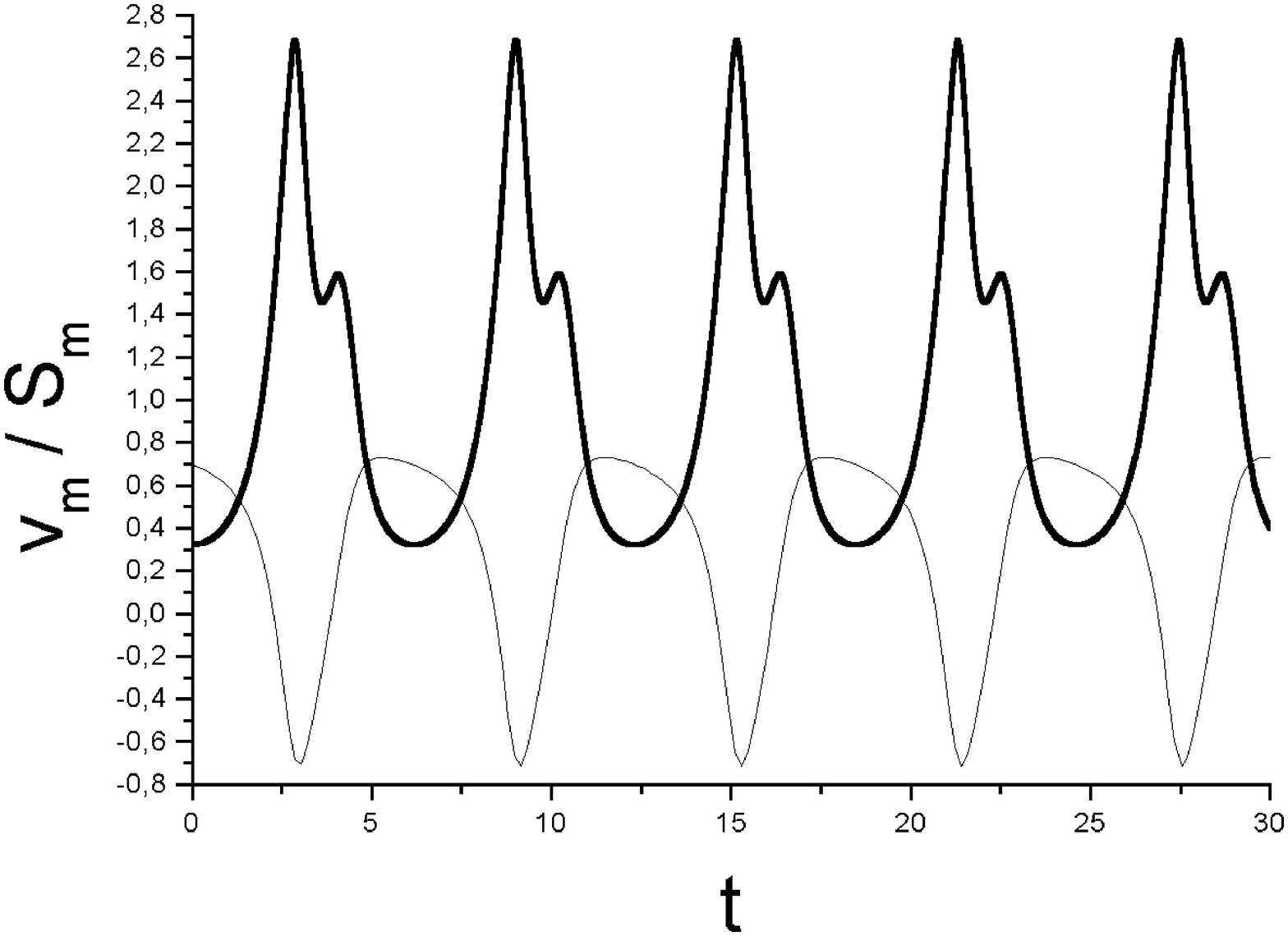}}}
\caption{} \label{fig:3}
\end{figure}

\newpage
\begin{figure}[h]
\centering
\resizebox{.5\columnwidth}{!}{\rotatebox{-90}{\includegraphics{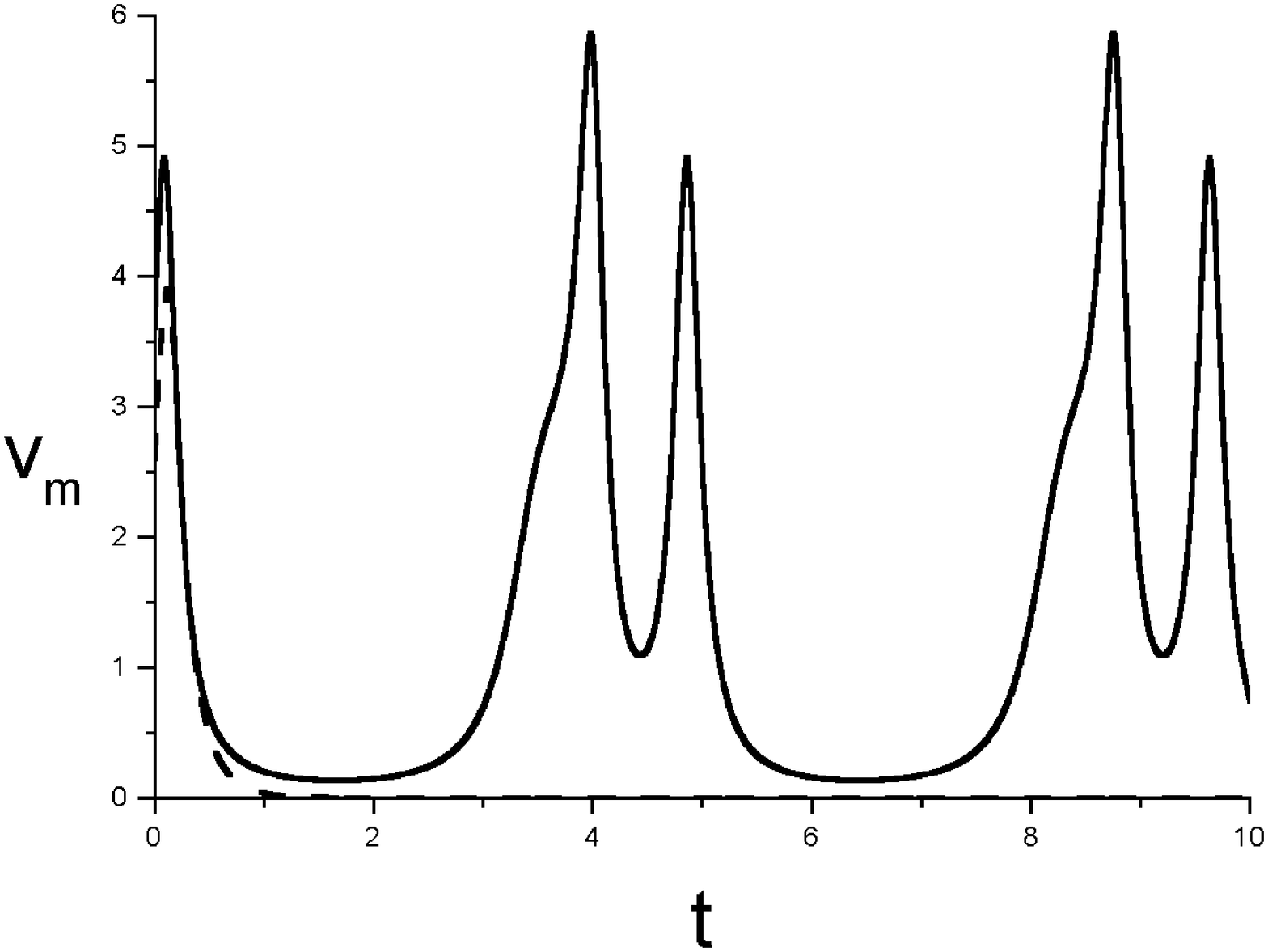}}}
\caption{} \label{fig:4}
\end{figure}

\newpage
\begin{figure}[h]
\centering
\resizebox{.5\columnwidth}{!}{\rotatebox{-90}{\includegraphics{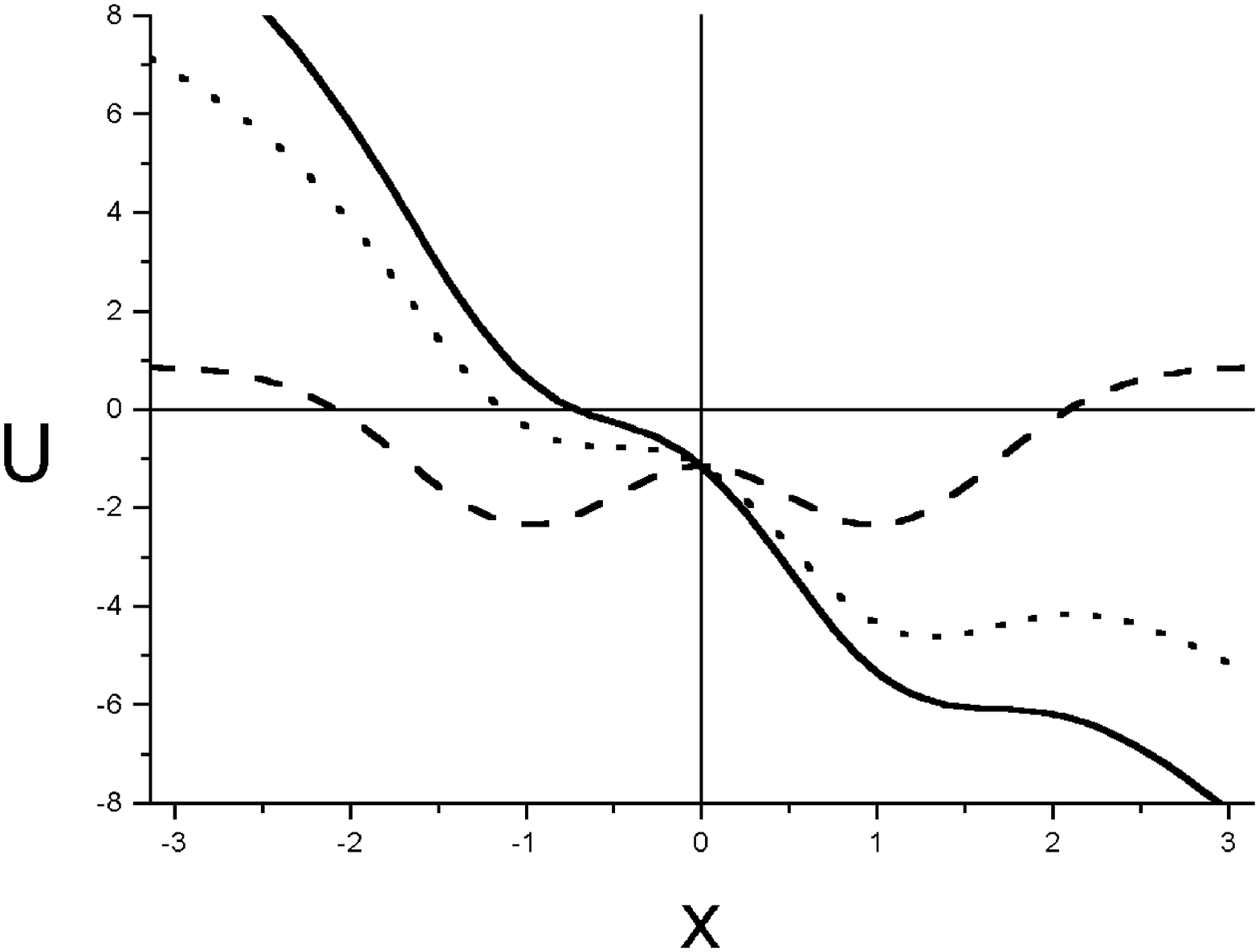}}}
\caption{} \label{fig:5}
\end{figure}

\end{document}